\documentclass[citeautoscript,floatfix,aps,prb,twocolumn,
      superscriptaddress]{revtex4-2}

\usepackage{amsmath}
\usepackage{graphicx}
\usepackage{epstopdf}
\usepackage{natbib}
\usepackage{array}
\usepackage[utf8]{inputenc}
\usepackage{xspace}
\usepackage{multirow}
\usepackage{changes}

\usepackage{dcolumn}
\usepackage{bm}
\usepackage{braket}
\usepackage{xr}
\usepackage{float}

\usepackage{soul}  

\newcommand\thefontsize{The current font size is: \f@size pt}

\newcommand{\cubic}{$Pm\overline{3}m$\xspace}
\newcommand{\smo}{SrMoO$_3$}
\newcommand{\swo}{SrWO$_3$}
\newcommand{\sto}{SrTaO$_3$}
\newcommand{\sno}{SrNbO$_3$}

\newcommand{\svo}{SrVO$_3$}
\newcommand{\prt}{$\rho(300\text{ K})$}

\newcommand{\runit}{$\mu\Omega\cdot\text{cm}$}

\RequirePackage[
  hyperindex,colorlinks,bookmarksnumbered,
  plainpages=true,pdfstartview=FitH]{hyperref}
\hypersetup{urlcolor={blue!50!black}, linkcolor={red!50!black},citecolor={blue!50!black}} 
\usepackage{hyperref}
\usepackage[all]{hypcap}

\begin{document}

\title{Electron-phonon origins of unconventional resistivity in moderately correlated perovskite oxides}

\author{Jennifer Coulter}
\email{jcoulter@flatironinstitute.org}
\affiliation{Center for Computational Quantum Physics,
             Flatiron Institute,
             162 5th Avenue, New York, New York 10010, USA.}
\author{Fabian B.~Kugler}
\affiliation{Institute for Theoretical Physics, University of Cologne, 50937 Cologne, Germany}
\affiliation{Center for Computational Quantum Physics,
             Flatiron Institute,
             162 5th Avenue, New York, New York 10010, USA.}
\author{Harrison LaBollita}
\affiliation{Center for Computational Quantum Physics,
             Flatiron Institute,
             162 5th Avenue, New York, New York 10010, USA.}
\author{Antoine Georges}
\affiliation{Center for Computational Quantum Physics,
             Flatiron Institute,
             162 5th Avenue, New York, New York 10010, USA.}
\affiliation{Coll{\`e}ge de France, 11 place Marcelin Berthelot, 75005 Paris, France}
\affiliation{CPHT, CNRS, {\'E}cole Polytechnique, IP Paris, F-91128 Palaiseau, France}
\affiliation{DQMP, Universit{\'e} de Gen{\`e}ve, 24 quai Ernest Ansermet, CH-1211 Gen{\`e}ve, Suisse}
\author{Cyrus E. Dreyer}
\email{cyrus.dreyer@stonybrook.edu}
\affiliation{Department of Physics and Astronomy,
             Stony Brook University,
             Stony Brook, New York, 11794-3800, USA}
\affiliation{Center for Computational Quantum Physics,
             Flatiron Institute,
             162 5th Avenue, New York, New York 10010, USA.}
\date{\today}

\begin{abstract}

Transition-metal perovskite oxides exhibit moderately correlated metallic phases,  several of which exhibit a $T^2$ resistivity scaling up to temperatures far exceeding the regime where Fermi-liquid electron-electron scattering is expected to dominate. Some of these materials, such as \smo{}, also exhibit unexplained ultra-low room-temperature resistivity.
We demonstrate that in \smo{}, \swo{}, \sto{}, \sno{}, and \svo{} electron-phonon scattering results in quadratic-scaling resistivity due to the shape of the Fermi surface and the thermal activation of optical phonons. We also reveal that the origin of the low resistivity of \smo{} is an overall low electron-phonon coupling strength, and identify \swo{} and \sto{} as other possible low-resistivity oxides. Additionally, we find that the strength of electron-phonon coupling is sensitive to structural distortions, energies of optical phonons, and the treatment of electronic correlations. This suggests design principles for finding other ultra-high conductivity transition-metal oxides, and has significant implications for theoretical interpretation of direct-current resistivity in transition-metal oxides and beyond. 

\end{abstract}

\maketitle

The temperature-dependent direct-current (DC) resistivity $\rho(T)$ is one of the most fundamental probes of the electronic properties of materials and is critical to the technological utility of metals. For example, materials with high electrical conductivity at room temperature (RT) and above are important in all electronic devices as contacts and interconnects that allow current to flow with minimal resistance, reducing power consumption and improving heat management \cite{Gall2020,Gall2021,Kim2024}. 

Transition-metal perovskite oxides, which can feature strong correlations and collective phenomena, exhibit some of the highest conductivities of all known metals \cite{Biswas2023,Rimal2025}. In particular, \smo{} has the eighth lowest reported RT resistivity~\cite{Nagai2005,Mackenzie2017}, lower than transition metals like platinum and palladium and alkali metals like lithium and potassium. As a result, \smo{}, \svo{}, \sno{}, and other materials in this family show promise as transparent conductors \cite{Zhang2016}. 

Yet, the origin of the ultra-low resistivity in these oxides remains an open question, in part because  $\rho(T)$ follows a $T^2$ scaling in an unusually wide range, as high as $\sim\! 150$ K in \smo{}~\cite{Nagai2005,Wang2001,Lekshmi2005,Radetinac2016,Cappelli2022}, $\sim\! 180$ K in \sno~\cite{Oka2015,Ok2021}, and even RT in \svo{}~\cite{Chamberland1971,Reyes2000,Inoue1998,berry2022,Ahn2022,xu2019,Roth2021,Fouchet2016,Brahlek2015,Brahlek2024,Shoham2020,Zhang2016,Mirjolet2019,mirjolet2021,Mirjolet2021_2}. Quadratic scaling is usually associated with Fermi-liquid (FL) electron-electron (el-el) scattering; however, el-el scattering is highly unlikely to determine transport in moderately correlated metals at such high $T$~\cite{Our_PRL}. Instead, in this regime, electron-phonon (el-ph) scattering is expected to dominate, although the canonical $T^5$ or $T$-linear scaling~\cite{Ziman1960} typically associated with el-ph scattering is not observed. To complicate matters further, there is significant spread in experimental measurements for $\rho(T)$, often with key differences between thin-film and single-crystal samples \cite{Our_PRL}. For example, the superlative conductivity of \smo{} comes from the only reported single-crystal sample~\cite{Nagai2005}, while for \svo{} the thin films show much lower resistivity than most of the single crystals at RT \cite{Our_PRL}.

In this work, we use \textit{ab-initio} el-ph coupling calculations based on density functional theory (DFT) in combination with the Boltzmann transport equation (BTE) to explain the DC resistivity in low-resistivity perovskite oxide metals. We demonstrate that the $T^2$ scaling arises from \emph{el-ph} scattering due to the shape of the Fermi surface~\cite{Kukkonen1978,Dylla2019,mirjolet2021} and the role of optical phonons~\cite{abramovitch2024}, and that the superlative conductivity of \smo{} is due to weak el-ph coupling (similar to, e.g., copper). We predict that several similar oxides (\swo{}, \sto{}, and \sno{}) could show similarly low RT resistivity, and show that the somewhat higher resistivity of \svo{} is explained via the role of electron correlations in the el-ph coupling of that material.

Finally, we comment on the role of structural distortions in achieving low RT resistivity, and how they may clarify the thin-film/single-crystal discrepancies in the literature. Trends uncovered in these calculations highlight design principles for the discovery of other ultra-low resistivity transition-metal oxides.

\textit{Methods} Transport calculations were performed by solving the linearized BTE as implemented in the Phoebe code~\cite{phoebe} by iterative solution~\cite{omini1995iterative}. To approximate the effects of Coulomb interactions within the transition-metal $d$ states at the Fermi level, we applied the simplified rotationally-invariant DFT+$U$ scheme of Dudarev et al.~\cite{dudarev1998electron} in which an effective $U_{\mathrm{eff}}= U-J$ is used; 
following prior work, we choose $U_{\rm eff} = 2.76$ eV for all materials \cite{LeeHand2020,Hampel2020,Cappelli2022,Liang2025} except \svo{}, where we use $3.85$ eV \cite{Lechermann2006}. Additional computational details are found in the Supplemental Material (SM)~\cite{SM}, Sec. S1. We note that the inclusion of $U$ does not significantly affect the electronic structure of these materials, but can influence the strength of the el-ph coupling as we show below. Recent work has underscored the connection between electron interactions and el-ph via more sophisticated methods than DFT+$U$ \cite{abramovitch2025, li2019, lazzeri2008, mandal2014}, but we expect the general trends to hold due to the relatively weak correlations in these materials. 

\begin{figure}
    \includegraphics[width=0.95\linewidth]{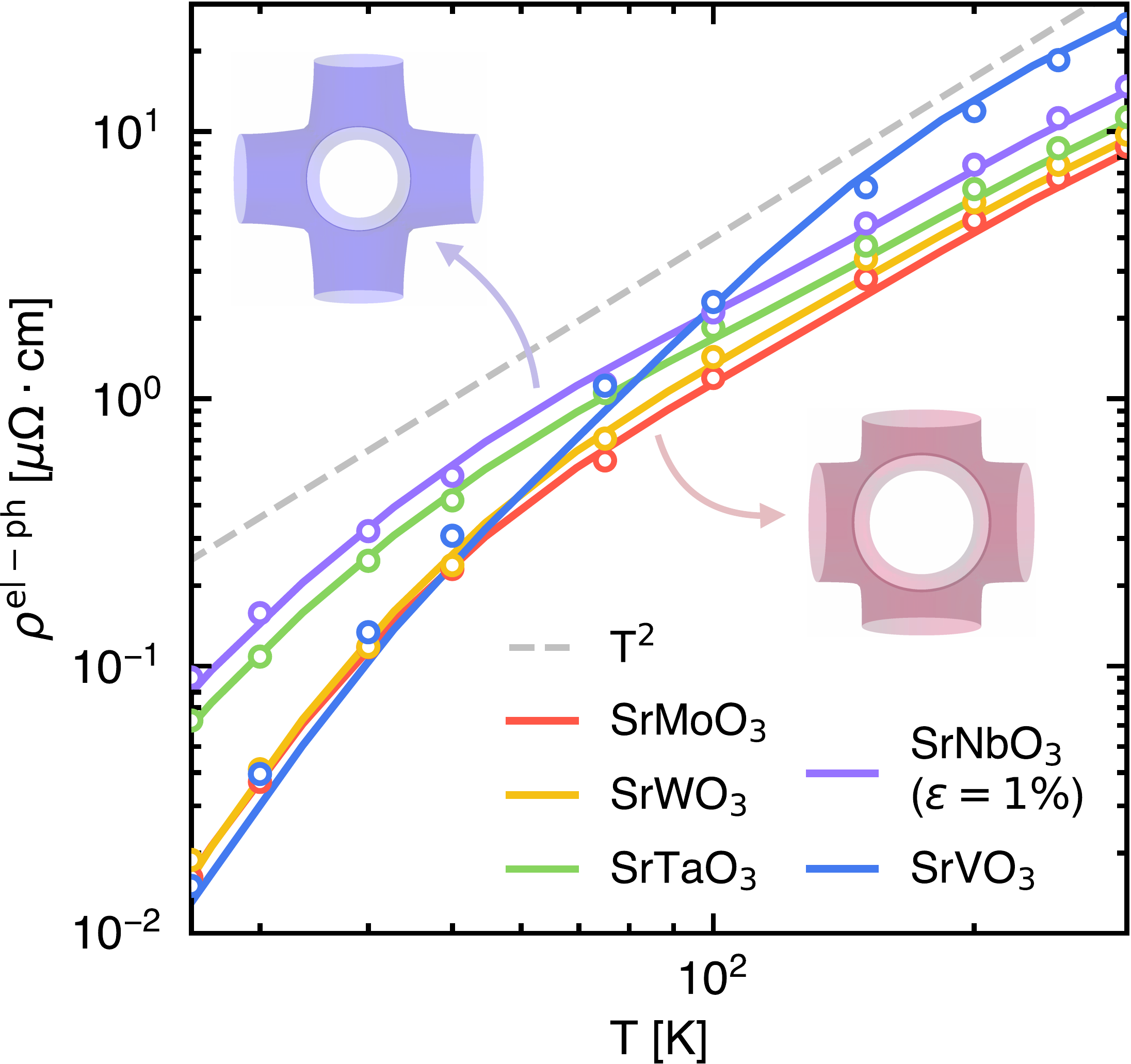}
    \caption{
    DC resistivity versus temperature of \cubic{} \smo{}, \swo, \sto, \sno (with 1\% tensile hydrostatic strain), and \svo, calculated using an iterative solution to the BTE from electron-phonon scattering (indicated by points) as well as fits to the Kukkonen model~\cite{Kukkonen1978} (indicated by solid lines). The gray dashed line demonstrates $T^2$ scaling, and the red (\smo{}) and purple (\sno{}) show the sheet of the Fermi surface which is comprised of interpenetrating cylinders.
    }
    \label{fig:resistivity}
\end{figure}

\textit{Temperature-dependent resistivities} Figure~\ref{fig:resistivity} shows the el-ph-limited resistivity for the cubic \cubic{} structure of the perovskite oxides \smo{}, \swo{}, \sto{}, \sno{}, and \svo{}, where points indicate the results of the BTE calculations and lines the fits to the Kukkonen model \cite{Kukkonen1978} (explained below).
All crystal structures were fully relaxed with the exception of \sno{}, which required a 1\% hydrostatic tensile strain to stabilize \cubic{}, indicated by the label $\varepsilon=1\%$.
We will first analyze the RT resistivities of the materials, explaining the observed trends and how to understand the varied experimental results of the resistivity in \smo{};
then, we will discuss the origin of the temperature scaling, especially the unconventional $T^2$ regime.

Consistent with experiment, we find a very low \prt{} for \smo{} of about 8.8 \runit{} which slightly \emph{overshoots} the experimental results of Ref.~\citenum{Nagai2005} (though only by $\sim4$ $\mu\Omega\cdot$cm). This could arise from theoretical uncertainties or experimental challenges of obtaining accurate aspect ratios to define $\rho$ in single crystals and/or thin films. Interestingly, \swo{} and \sto{}, whose $\rho(T)$ has not been experimentally characterized, also show a low RT value. In our calculations, \svo{} has the largest RT resistivity; there are still significant open questions about the experimental situation~\cite{Reyes2000,Inoue1998,berry2022,Ahn2022,xu2019,Roth2021,Fouchet2016,Brahlek2015,Brahlek2024,Shoham2020,Zhang2016,Mirjolet2019,mirjolet2021,Mirjolet2021_2,Our_PRL}, but the lowest reported value of 20 \runit{}~\cite{Brahlek2024} at RT agrees well with our value of $\sim$25 \runit{}. Finally, for \sno{} we find a somewhat smaller RT resistivity than recent measurements \cite{Oka2015,Ok2021,Kumar2024,Palakkal2025} as discussed below. 

To understand the magnitudes and trends in the el-ph limited RT resistivities, we consider a simplified isotropic version of the Bloch–Gr\"{u}neisen (BG) equation (applicable at high $T$)~\cite{allen1978_phys_today_BG}, $\rho_{\text{BG}}(T) \approx 6\pi V k_B T \lambda_{\rm tr}/e^2 \hbar N_{\text{F}} \langle v^2 \rangle_{\rm FS} $, where $V$ is the unit-cell volume, $\langle v^2 \rangle_{\rm FS}$ is the square of the band velocity averaged over the Fermi surface, $N_{\text{F}}$ is the electronic density of states (DOS) at the Fermi level, and $\lambda_{\rm tr}$ is the cumulative isotropic transport el-ph coupling strength~\cite{grimvall1999thermophysical, ponce2016epw}. The latter is given by
\begin{multline}
\lambda_{\mathrm{tr}} = \frac{1}{N_{\mathrm{F}}} \sum_{m n, \nu} 
\int_{\rm BZ} 
\frac{d \mathbf{q}}{V_{\text{BZ}}} 
\int_{\rm BZ}  
\frac{d \mathbf{k}}{V_{\text{BZ}}} 
\\ \times \frac{1}{\omega_{\mathbf{q} \nu}} \left|g_{m n, \nu} (\mathbf{k}, \mathbf{q})\right|^2 \left( 1 - \frac{v_{\textbf{k}m}\cdot v_{\textbf{k+q}n}}{|v_{\textbf{k}m}||v_{\textbf{k+q}n}|} \right) \\ \times\delta\left(\epsilon_{\mathbf{k}m} - \epsilon_{\text{F}} \right) \delta\left(\epsilon_{\mathbf{k}+\mathbf{q}n} - \epsilon_{\text{F}} \right), 
\label{eq:lambda}
\end{multline}
where $v_{\textbf{k}m}$ is the velocity of band $m$ and $k$-point \textbf{k}, $\omega_{\mathbf{q} \nu}$ is the frequency of the phonon branch $\nu$ and wavevector \textbf{q}, $V_{\text{BZ}}$ is the volume of the Brillouin Zone (BZ), and $\epsilon_{\text{F}}$ the Fermi energy. $g_{m n,\nu}(\mathbf{k}, \mathbf{q})$ are the electron-phonon matrix elements,
\begin{equation}
g_{m n,\nu}(\mathbf{k}, \mathbf{q})=\left(\frac{\hbar}{2 m_0 \omega_{\mathbf{q} \nu}}\right)^{1 / 2} \left\langle\psi_{\mathbf{k}+\mathbf{q},m}\right| \partial_{\mathbf{q} \nu} V\left|\psi_{\mathbf{k}n}\right\rangle ,
\label{eq:g}
\end{equation}
where $\psi_{\textbf{k}n}$, $\psi_{\textbf{k}+\textbf{q},m}$ are the Kohn--Sham (KS) wavefunctions at bands $n,m$ and wavevectors \textbf{k}, $\textbf{k}+\textbf{q}$, $m_0$ a reference mass~\cite{giustino2007}, and $\partial_{\mathbf{q} \nu} V$ the change in KS potential with the phonon displacement.

\begin{table}[h!]
\begin{center}
\begin{ruledtabular}
\begin{tabular}{c|cccc}
& $\langle v^2 \rangle_{\mathrm{FS}}$ & $\frac{N_{\text{F}}}{V}$ & $\lambda_{\rm tr}$  & $\rho$(300 K) \\
 & $\left[10^{12}\frac{m^2}{s^2}\right]$ & $\left[\frac{\rm eV}{\text{\AA{}}^3}\right]$ & & [$\mu \Omega \cdot $ cm] \\[1ex]
\hline
Cu & 1.47 & 0.024 &  0.12 & 1.47   \\ 
\smo{} & 0.43                   & 0.024 & 0.14  & 8.79 \\
\swo{}      & 0.51                   & 0.026 & 0.13 & 9.74   \\ 
\sto{}      & 0.51                   & 0.017 & 0.11 & 11.34 \\
\sno{} $\varepsilon=1\%$     & 0.41                   & 0.017  & 0.12 & 14.77  \\
\svo{}     & 0.21                   & 0.028  & 0.19 & 25.25 \\
$Imma$ \smo{} & 0.26 & 0.028 & 0.17 & 31.18  \\ 
$Pmna$ \sno{} & 0.28                   & 0.017           & 0.12  & 17.83 
\end{tabular}
\end{ruledtabular}
\end{center}
 \caption{Calculated squared Fermi velocity averaged over the Fermi surface, density of states at the Fermi level, and isotropic transport el-ph coupling strength for Cu, \smo{} (cubic and orthorhombic), \swo{}, \sto{}, \sno{} (cubic with 1\% tensile hydrostatic strain and orthorhombic), and \svo{}.  }
 \label{table:lambda_parameters}
\end{table}

\textit{Room-temperature resistivity analysis} In Table~\ref{table:lambda_parameters}, we provide our results for $\rho(300\text{ K})$ together with the band velocities, DOS, and $\lambda_{\mathrm{tr}}$ for the perovskite materials. For reference, we also include copper metal. 
The cubic perovskite oxides, with the exception of \svo{}, have relatively weak el-ph interactions, with $\lambda_{\mathrm{tr}}$ similar to Cu. 
%
The key difference is that Cu has a much larger band velocity; $\langle v^2 \rangle_{\mathrm{FS}}$ is similar across the cubic oxides, though as expected, increases as the transition-metal atom moves down the periodic table.
%
The $d^2$ oxides \smo{} and \swo{} have similar DOS at the Fermi level as Cu, while $d^1$ \sto{} and \sno{} have a somewhat smaller values (this is not the case for \svo{} due to the significantly smaller bandwidth). 

Table~\ref{table:lambda_parameters} also includes the orthorhombic structures of \smo{} ($Imma$) and \sno{} ($Pmna$). As expected, the structural distortions significantly reduce the bandwidth and thus decrease the Fermi velocity (leaving the volume-normalized DOS relatively unchanged). In \sno{}, this yields a modest increase in \prt{}, closer to the experimental measurements \cite{Oka2015,Ok2021,Kumar2024,Palakkal2025}. For \smo{}, we see a much larger increase in \prt{} with the distortions to $Imma$, driven by a significant increase in the strength of the el-ph coupling.

\begin{figure*}[hbtp]
    \centering
    \includegraphics[width=1.0\textwidth]{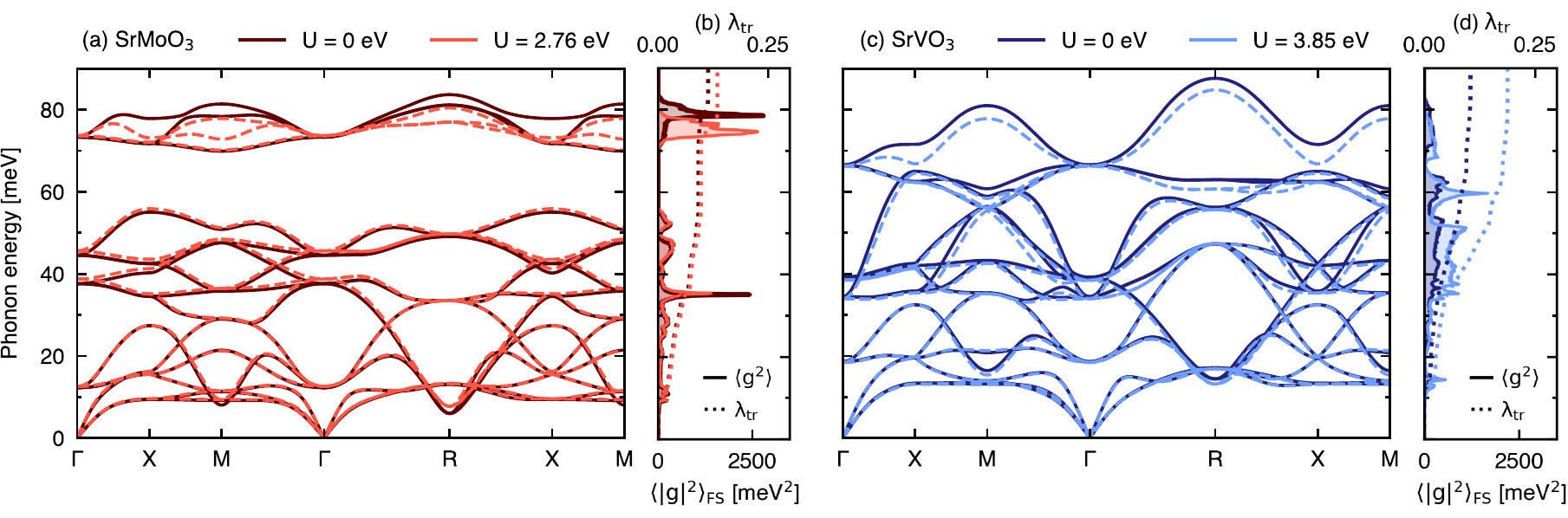}
    \caption{Phonon dispersion for (a) \smo{} and (c) \svo{} with and without Hubbard $U$. Panels (b) and (d) show the el-ph matrix elements averaged over the Fermi surface  [Eq.~\eqref{eq:g_fermi}] (solid curves) and isotropic transport el-ph coupling strength $\lambda_{\text{tr}}$ [Eq.~\eqref{eq:lambda}] resolved over phonon frequencies (dashed curves) versus phonon energy for \smo{} and \svo{}, respectively.}
    \label{fig:phonons+elph}
\end{figure*}

The larger $\lambda_{\mathrm{tr}}$ in \svo{} and $Imma$ \smo{} compared to the other materials in Table~\ref{table:lambda_parameters} is understood by examining the phonon dispersions of these materials and the effect of electron correlations on the el-ph coupling. In Fig.~\ref{fig:phonons+elph}, we compare the phonon dispersions of \cubic{} \smo{} and \svo{} (see SM~\cite{SM} Figs. S3 and S2 for the other oxides). In panels (b) and (d) we show $\lambda_{\text{tr}}$ versus phonon frequency, and the el-ph matrix elements averaged over the Fermi surface, given by
\begin{multline}
\langle |g|^2 \rangle_{\mathrm{FS}}(\omega) = \sum_{m n, \nu} 
\int_{\rm BZ}  
\frac{d \mathbf{q}}{V_{\text{BZ}}} 
\int_{\rm BZ}  
\frac{d \mathbf{k}}{V_{\text{BZ}}} 
\left|g_{m n, \nu}(\mathbf{k}, \mathbf{q})\right|^2 \\ \times\delta\left(\epsilon_{n \mathbf{k}} - \epsilon_{\text{F}}\right) \delta\left(\epsilon_{m \mathbf{k}+\mathbf{q}} - \epsilon_{\text{F}}\right) \delta(\omega - \omega_{\mathbf{q}\nu}).
 \label{eq:g_fermi}
\end{multline}
Results with and without Hubbard $U$ are included in Fig.~\ref{fig:phonons+elph}.

We find that the modes with the strongest el-ph coupling in \smo{} are the high-frequency optical modes which involve primarily O displacements (see SM~\cite{SM} Fig. S4). These are also the modes that are most affected by the inclusion of Hubbard $U$. Generally, a strong effect of $U$ is not expected for nonmagnetic \cubic{} perovskites; however, we show in SM~\cite{SM} Fig. S4 that the high-frequency modes couple most strongly to splittings of the $t_{2g}$ bands, and thus the treatment of the Coulomb interactions plays a key role. We see in Fig.~\ref{fig:phonons+elph} that these modes are much lower in energy in \svo{}, and thus their enhancement with $U$ has a larger effect on transport around RT than in \cubic \smo{}. For $Imma$ \smo{}, on the other hand, these modes are pushed significantly lower by the structural distortions, which is the reason for the enhanced el-ph scattering; this shift does not occur for $Pnma$ \sno{} (SM~\cite{SM} Fig. S3).

\begin{figure}
    \centering
    \includegraphics[width=0.95\linewidth]{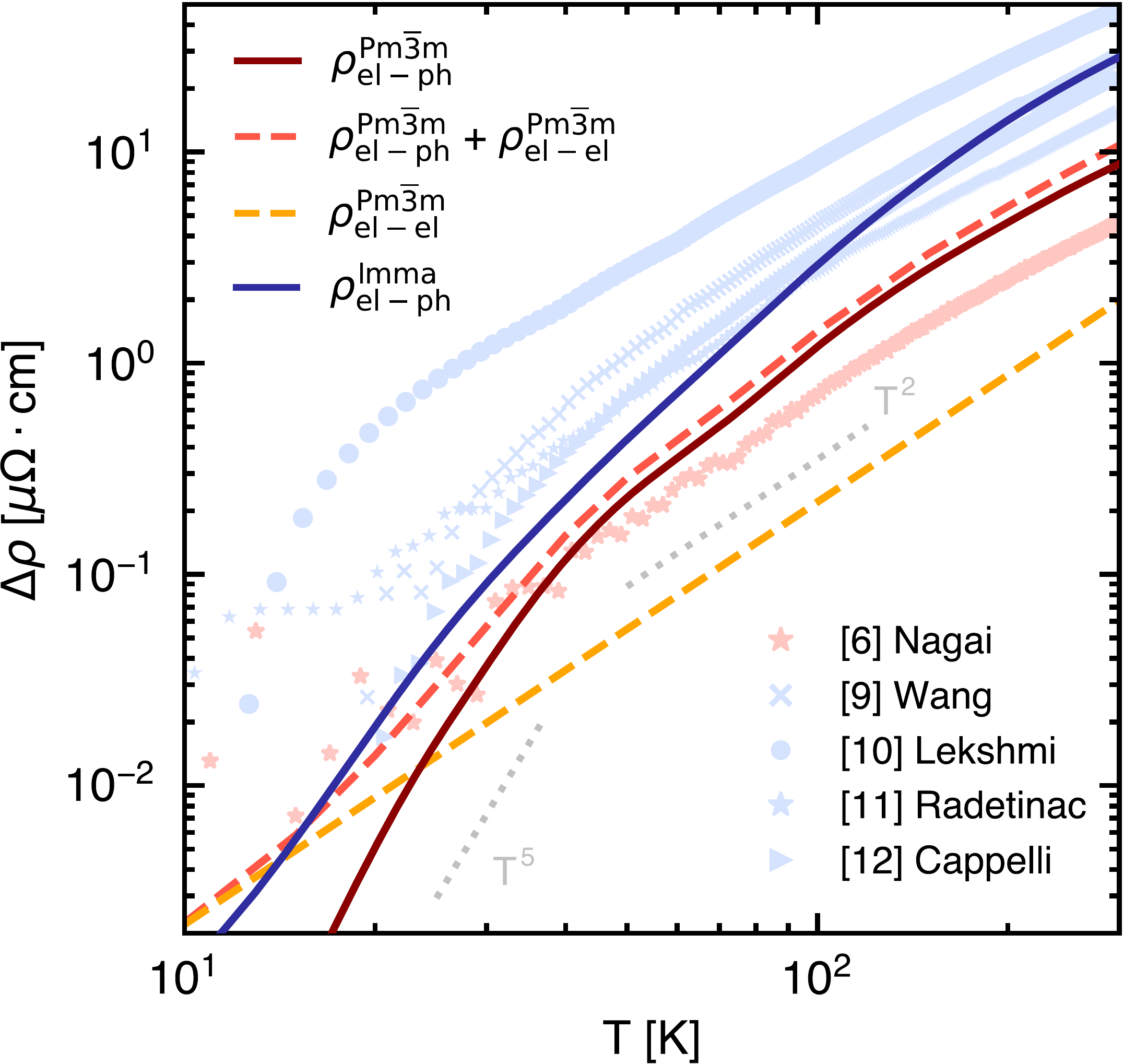}
    \caption{
    DC resistivity versus temperature for \smo{}.
    The solid red and blue lines are el-ph resistivities from iterative BTE solutions using DFT+$U$ data for the $Pm\overline{3}m$ and $Imma$ structures, respectively.
    The red dashed line further contains the el-el resistivity for the $Pm\overline{3}m$ structure, taken from Ref.~\citenum{Our_PRL} and separately shown in yellow.
    Light blue points are experimental measurements for thin films~\cite{Wang2001,Lekshmi2005,Cappelli2022,Radetinac2016}, and light red for the single crystal~\cite{Nagai2005}. Residual resistivities are subtracted as in Ref.~\citenum{Our_PRL}.
    Gray dotted curves are a guide to the eye indicating $T^5$ and $T^2$ dependencies.
    \label{fig:smo_resistivity}}
\end{figure}

\textit{The case of \smo{}} We can use these results to understand the experimental situation of \smo{}. 
In Fig.~\ref{fig:smo_resistivity}, we compare our calculations for \smo{} to experimentally reported resistivities, including data reproduced from the single-crystal measurements of Ref.~\citenum{Nagai2005} and the thin-film measurements of Refs.~\citenum{Wang2001,Lekshmi2005,Radetinac2016,Cappelli2022}. In all cases, we removed the residual resistivities as in Ref.~\citenum{Our_PRL} to obtain $\Delta\rho(T)$. For the cubic case, we also include the contribution from el-el scattering determined from DFT plus dynamical mean-field theory (DMFT)~\cite{Our_PRL}, and the total resistivity, for which we find Matthiessen's rule to be a good approximation~\footnote{In this instance, adding the DMFT Fermi-liquid el-el scattering rate to the el-ph rate within the relaxation-time approximation produces the same result as adding $\rho_{\mathrm{DMFT}}$ and $\rho_{\rm el-ph}$ calculated separately using the RTA level. However, in the main manuscript, we use an iterative solution to the linearized BTE, which takes into account the full electron scattering matrix in the solution of the BTE. Here, our DMFT el-el scattering rate only provides momentum-independent ``out'' scattering, representing the diagonal contribution to the scattering matrix, rather than a full scattering-matrix contribution. To avoid violating conservation of charge in the full scattering-matrix BTE calculation, we calculate and add these contributions to $\rho$ separately.} and treat $\rho_{\mathrm{el-el}}$ and $\rho_{\mathrm{el-ph}}$ as additive. 

The el-ph contribution to $\Delta\rho(300 \text{K})$ is a factor of $\sim 4$ larger than the el-el contribution and dominates the $T$ dependence down to $\sim 40$ K. Then, as the resistivity from el-ph scattering transitions from $T^2$ to $T^5$, el-el scattering becomes relevant and FL el-el $T^2$ behavior appears around $\sim10$ K. This clearly indicates that the $T^2$ scaling above $\sim40$ arises from el-ph scattering. The same is true for \svo{}, whose FL el-el scattering behavior is very similar to \smo{} \cite{Our_PRL}. 

In addition, the calculated difference in resistivity between \cubic{} and $Imma$ \smo{} is roughly the difference in $\Delta\rho(300\text{K})$ between the thin-film and the single-crystal experimental results (though there is considerable spread of the data in the former). Based on neutron diffraction of powders, the cubic structure is expected to be the ground state at RT~\cite{Macquart2010}; however, some of the thin films~\cite{Radetinac2016,Cappelli2022} were grown on orthorhombic substrates which could stabilize $Imma$ at RT (Ref.~\citenum{Cappelli2022} explicitly reported the $Imma$ structure). Even on cubic substrates, unrelaxed lattice mismatch will cause distortions away from cubic. In principle, the single crystal~\cite{Nagai2005} has no such constraints and is expected to be cubic. A similar observation has been made for \sno{}, where relaxed films show lower resistivity than strained ones \cite{Ok2021,Rimal2025}. Therefore, we believe that structural distortions may play a significant role in the discrepancy between single crystals and thin films. There may be other differences in properties (e.g., structural defect densities, impurity type and concentration, oxygen stoichiometry) for different synthesis methods, but direct simulation of these effects is beyond the scope of current methods for electrical conductivity predictions.

\textit{Origin of $T^2$ resistivity scaling} Finally, we turn to the origin of the unexpected $\sim T^2$ behavior from el-ph scattering observed in all of the oxides (see Fig.~\ref{fig:resistivity}). A key element to explaining this behavior is the fact that the Fermi surface in the $t_{2g}$ perovskite oxides under study has cylindrical sheets~\cite{mirjolet2021,Cappelli2022}, as shown in the insets in Fig.~\ref{fig:resistivity}. As discussed by Kukkonen~\cite{Kukkonen1978} and in Ref.~\citenum{Dylla2019} (see End Matter for details), a cylindrical Fermi surface results in three distinct temperature regimes, determined by the dimensions of the cylinder, i.e., its diameter $2k_r$ and length $2k_l$. These define temperature scales $T_d=\hbar s 2 k_r/k_{\text{B}}$ and $T_l=\hbar s  2k_l/k_{\text{B}}$, respectively ($s$ is an isotropic speed of sound in the material).  For very low $T$ ($T \ll T_d$), $\rho$ is expected to follow a $T^5$ dependence based on the normal BG arguments \cite{Ziman1960,Kukkonen1978}. Similarly, for very high $T$ ($T\gg T_l,\Theta_{\text{D}}$, where $\Theta_{\text{D}}$ is the Debye temperature), $\rho$ will be proportional to $T$ since, from BG, the only $T$ dependence comes from the phonon distribution function, which is linear in $T$. 

The interesting regime is $T_d\lesssim T \lesssim T_l$. Here, large-angle scattering dominates in the radial direction of the cylinder. Assuming deformation-potential-like scattering matrix elements ($\propto q$), the phase space for scattering along the length of the cylinder increases with $T$. Finally, the phonon distribution function is also linear in $T$, as it depends on the intersection between the phonon sphere (assuming a Debye phonon) of radius $q_T=k_{\text{B}}/\hbar s$ and the cylindrical Fermi surface, and thus is approximately $\pi k_r^2 2q_T\propto T$. Therefore, in total, we have $\rho(T)\propto T^2$.

To demonstrate the correspondence of this model and our calculation, we fit our DFT+BTE el-ph resistivity for the $Pm\overline{3}m$ structures to the model of Ref.~\citenum{Kukkonen1978}. The fits are presented as lines in Fig.~\ref{fig:resistivity} (see Table~\ref{table:Kuk_fits} of the End Matter for the fitting parameters). In addition to the acoustic (Debye) mode, we must include an additional optical (Einstein) mode in order to obtain a good fit. In Fig.~\ref{fig:rho_kukkonen} of the End Matter we show the contributions of the different phonon modes to the total resistivity and also compare to the Bloch model, which assumes a spherical Fermi surface. For all oxides other than \svo{}, the Kukkonen model provides a superior fit compared to the Bloch model, capturing the $T^2$ regime at intermediate $T$. Importantly, the onset of the $T^2$ regime (cf.\ the shaded blue regions in Fig.~\ref{fig:rho_kukkonen}) scales with the diameter of the Fermi-surface cylinder, see Fig.~\ref{fig:rho_kukkonen}(f). The $d^1$ materials \sto{} and \sno{} have a smaller diameter, and thus the $T^2$ regime is shifted to lower $T$ compared to the $d^2$ materials \smo{} and \swo{}. For these materials (apart from \svo{}), the Einstein mode does not affect $T^2$ scaling, only the high-$T$ $T$-linear regime.

As before, the odd material out is \svo{}. In this case, the $T^2$ regime occurs at higher $T$, even though it is $d^1$ and has a narrow cylinder. We find that the $T$ dependence is here determined by the low-energy optical modes, as argued by Ref.~\citenum{abramovitch2024}. Indeed, to get a good fit, we must take the energy of the Einstein mode to be 50 meV (as opposed to 70 meV for the rest of the oxides, see Table~\ref{table:Kuk_fits}), and for the Kukkonen model we must even consider an additional optical mode at 20 meV. Then, from Fig.~\ref{fig:rho_kukkonen}(e), the Kukkonen and Bloch models fit the data equally well, indicating that the Fermi surface geometry is not the main contributor. This confirms the claim in Ref.~\citenum{abramovitch2024} that in \svo{}, the thermal activation of optical modes is responsible for the apparent $T^2$ behavior. 

\textit{Conclusion} This work demonstrates that the superlative conductivity of a family of cubic perovskite-oxide metals arises from weak el-ph scattering, which dominates its resistivity at RT over FL el-el scattering. The unexpected $T^2$ dependence at intermediate $T$ in these materials is explained by the cylindrical shape of the Fermi surface and, as concluded in Ref.~\cite{abramovitch2024}, the thermal activation of optical modes.

This serves as a word of caution when interpreting high-$T$ $T^2$ behavior of the resistivity, which is observed in several other systems. These results also suggest two key design principles for finding other high-conductivity oxides: (i)\emph{stiff optical modes:} O-related optical phonons which have the largest contribution to el-ph scattering (especially when electron correlations are present) should be at high energy to minimize their effect on RT transport; and (ii) \emph{high symmetry:} distortions away from \cubic{} decrease the band width and can also soften the high-frequency optical modes, thus increasing the resistivity. Finally, the comparison between \svo{} and the other materials indicates the importance of electron correlation effects, motivating further developments in beyond-DFT \textit{ab-initio} el-ph coupling calculations.

\acknowledgements
We are grateful to Andrew Millis, Jernej Mravlje, Sophie Beck, David Abramovitch, and Philip Allen for useful discussions.
We thank Naoki Shirakawa for sending us the data of Ref.~\citenum{Nagai2005}.
C.E.D.\ acknowledges support from the National Science Foundation under Grant No.~DMR-2237674.
F.B.K.\ acknowledges funding from the Ministerium f\"ur Kultur und Wissenschaft des Landes Nordrhein-Westfalen (NRW-R\"uckkehrprogramm).
The Flatiron Institute is a division of the Simons Foundation. 

\bibliography{bibfile}

\clearpage

\section*{End Matter: $T^2$ resistivity from electron-phonon scattering on a cylindrical Fermi surface}
\label{sec:kukkonen_model}

As discussed in the main text, the most striking feature of the experimental and theoretical resistivity is the approximate $T^2$ dependence arising from el-ph scattering. To demonstrate how this can be caused by a cylindrical Fermi surface, we compare in this section our full DFT+BTE calculations to the the model of Kukkonen \cite{Kukkonen1978}, which assumes a cylindrical Fermi surface, as well as to the conventional Bloch–Gr\"{u}neisen equation~\cite{Ziman1960}, which assumes a spherical Fermi surface. The general argument is as follows. We can write the resistivity as 
\begin{equation}
\rho=\langle |g(\textbf{k},\textbf{q})|^2  n(\textbf{q}) (1-\cos\theta)\rangle ,
\end{equation}
where $g(\textbf{k},\textbf{q})$ is the el-ph matrix element, $n(\textbf{q})$ is the phonon distribution function, and the $1-\cos\theta$ factor isolates the high-angle scattering that degrades the current. We assume an isotropic Debye model for phonons characterized by a speed of sound $s$ and Debye temperature $\Theta_{\text{D}}=\hbar s Q_{\text{D}}/k_{\text{B}}$ ($Q_{\text{D}}$ is the Debye wavevector). The scattering is taken to be deformation-potential-like, $|g(\textbf{k},\textbf{q})|^2\propto q$ \cite{Ziman1960}, and the Fermi surface is a single cylindrical sheet with a diameter in reciprocal space of $2k_r$ and a length $2k_l > 2k_r$. Thus, we can define three temperature regimes. The first is $T \ll T_r, \Theta_{\text{D}}$ where $T_r=\hbar s 2k_r/k_{\text{B}}$; in this case, the maximum wavevector of the available phonons scales like $T$, so all $q$ dependence becomes $T$ dependence. Specifically, $|g(\textbf{k},\textbf{q})|^2\propto T$ and $(1-\cos\theta)\propto T^2$ \cite{Ziman1960}. The phonons that are allowed by energy and momentum to scatter lie on the intersection of the Fermi surface and the phonon sphere (the latter with radius $q_T=k_{\text{B}}/\hbar s$). For $q_T \ll k_r \ll k_l$, this intersection is a disk of area $\pi q_T^2$, and thus  $n(\textbf{q}) \propto T^2$. Putting this all together gives the conventional $T^5$ scaling. At very high temperatures $T \gg \Theta_{\text{D}}, T_r, T_l$, the only $T$ dependence is from the distribution function, which is linear in $T$. 

The interesting $T$ regime is $T_r\lesssim T \lesssim T_l$. In this case, there are sufficient phonon wavevectors for large-angle scattering across the cylinder, so $(1-\cos\theta)$ does not give any $T$ dependence. However, $|g(\textbf{k},\textbf{q})|^2$ still contributes a factor of $T$ since increasing $q$ allows more scattering along the length of the cylinder. Finally, the intersection with the phonon sphere and the Fermi surface is the cylindrical surface $\pi k_r^2 2q_T\propto T$ (neglecting the ends of the cylinder). Thus, the total dependence is $T^2$.  
\begin{figure*}
    \centering
    \includegraphics[width=\linewidth]{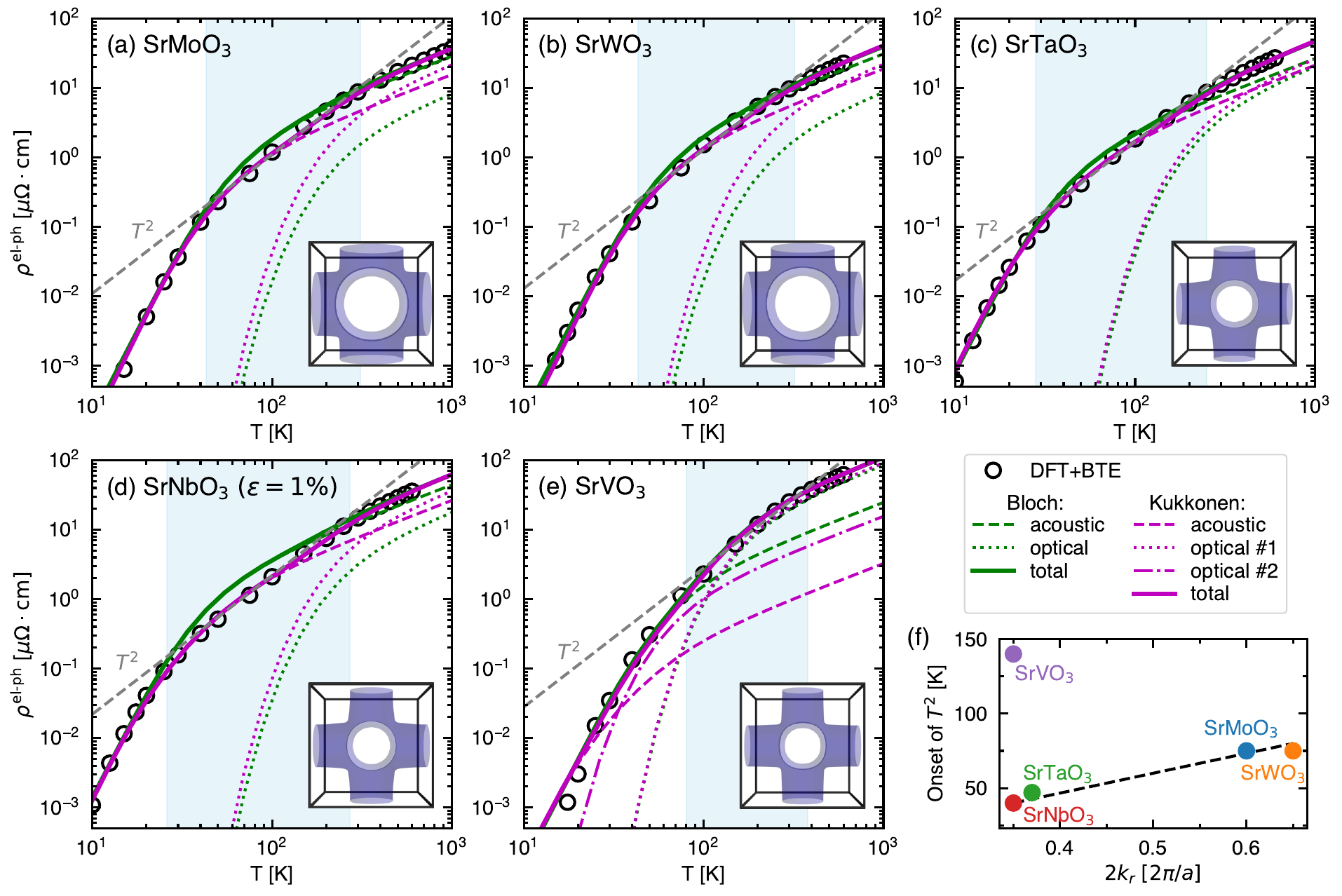}
    \caption{\textit{Ab-initio} DFT+$U$+BTE temperature-dependent el-ph resistivities (black circles) fit to the model of Ref.~\citenum{Kukkonen1978}, Eq.~(\ref{eq:kukkonen}), (magenta solid curves) which assumes a cylindrical Fermi surface, and the conventional Bloch–Gr\"{u}neisen formula, Eq.~(\ref{eq:BG}) (green solid curves), which assumes a spherical Fermi surface. Dashed (dotted) magenta and green curves are contributions from Debye acoustic (Einstein optical) modes to the Kukkonen and Bloch fits, respectively. For \svo{}, the Kukkonen fit required an additional optical mode indicated by the dot-dash curve in panel (e). Gray lines are guides to the eye for $T^2$ scalings and blue shaded regions indicate the temperature range of $T^2$ scaling. Panel (f) shows the onset of the el-ph $T^2$ regime as a function of the diameter of the Fermi-surface cylinder.
    \label{fig:rho_kukkonen}}
\end{figure*}

\begin{table*}
\begin{center}
\begin{ruledtabular}
\begin{tabular}{c|ccccccccc}
& $a$ [\AA{}] & $m^*$ [$m_e$] & $n$ [$e$/u.c.] & $\Theta_{\text{D}}$ [K] & s [m/s] & $E_{\text{Ein}} [eV]$ & $2k_r$ $[2\pi/a]$ & $C_{\text{Deb}}$ [eV] & $C_{\text{Ein}}$ [eV] \\\hline
\smo{}& 3.998 & 1.25 & 2 & 25.9 $( 25.9 )$ & 4999 $( 4999 )$ & 0.07 & 0.60 & 45 $( 33 )$ & 205 $( 60)$ \\
\swo{}& 4.013 & 1.0 & 2 & 25.9 $( 25.9 )$ & 5016 $( 5016 )$ & 0.07 & 0.65 & 65 $( 50 )$ & 265 $( 90 )$ \\
\sto{}& 4.07 & 1.0 & 1 & 25.9 $( 17.2 )$ & 5088 $( 3392 )$ & 0.07 & 0.37 & 75 $( 15 )$ & 365 $( 67 )$ \\
\sno{} $\varepsilon=1\%$& 4.11 & 1.25 & 1 & 25.9 $( 17.2 )$ & 5138 $( 3425 )$ & 0.07 & 0.35 & 60 $( 13 )$ & 310 $( 42 )$ \\
\svo{}& 3.873 & 2.0 & 1 & 25.9 $( 25.9 )$ & 4843 $( 4843 )$ & 0.05, 0.02 & 0.35 & 13 $( 9 )$ & 215, 35 $( 40 )$ \\
\end{tabular}
\end{ruledtabular}
\end{center}
 \caption{Parameters for the fits to the Kukkonen \cite{Kukkonen1978} and Bloch–Gr\"{u}neisen \cite{Ziman1960} models [Eqs.~(\ref{eq:kukkonen}) and (\ref{eq:BG})] given in Fig.~\ref{fig:rho_kukkonen}. $a$ is the lattice constant, $m^*$ the effective mass, $n$ the carrier density, $\Theta_{\text{D}}$ the Debye temperature of the Debye mode, $E_{\text{Ein}}$ the energy of the Einstein mode, $2k_r$ the diameter of the Fermi-surface cylinder, and $C_{\text{Deb}}$ ($C_{\text{Ein}}$) are the coupling constants for the Debye (Einstein) mode(s). Quantities in parentheses are for the Bloch–Gr\"{u}neisen model. Note that for \svo{}, two Einstein modes are used in the fit to the Kukkonen model. }
 \label{table:Kuk_fits}
\end{table*}

In Ref.~\citenum{Kukkonen1978}, the following expression for the resistivity was derived using the variational method that leads to the conventional Bloch--Gr\"{u}neisen equation (see below)~\cite{Ziman1960} but assuming a cylindrical Fermi surface (and neglecting Umklapp processes):

\begin{multline}
\label{eq:kukkonen}
    \rho(T)=\frac{9(m^*)^2C^2}{64\pi^4\hbar e^2 M n^2k_{\text{B}}T}\int_0^{2k_r}\frac{q_\perp^2 dq_\perp}{\sqrt{1-(q_\perp/2k_r)^2}}\times \\ \int_{q_\perp}^{q_{\text{max}}}\frac{q^3 dq}{\sqrt{q^2-q_\perp^2}(1-e^{-\hbar\omega(q)/k_{\text{B}}T})(e^{\hbar\omega(q)/k_{\text{B}}T}-1)},
\end{multline}

where $m^*$ is the effective mass, $C$ is the deformation potential, $M$ is the mass density, $n$ is the charge density, $\omega(q)$ is the phonon dispersion, and $q_{\text{max}}$ is the lesser of $Q_{\text{D}}$ and $\sqrt{(2k_l)^2+q_{\perp}^2}$ (for all materials, $2k_l=2\pi/a$). Note the minus sign in the exponent of the first exponential factor, which is missing in Ref.~\citenum{Kukkonen1978} (we attribute this to a typo in that work). We also compare with a fit to the conventional Bloch--Gr\"{u}neisen formula which assumes a spherical Fermi surface. Then, the resistivity is given by \cite{Ziman1960}
\begin{multline}
\label{eq:BG}
    \rho(T)=\frac{(m^*)^2C^2}{12\pi^3\hbar e^2Mn^2k_{\text{B}}T } \times \\ \int_0^{Q_{\text{D}}}\frac{q^5 dq}{(1-e^{-\hbar\omega(q)/k_{\text{B}}T})(e^{\hbar\omega(q)/k_{\text{B}}T}-1)}.
\end{multline}

The parameters for the fits are given in Table~\ref{table:Kuk_fits}. For \smo{}, \swo{}, \sto{}, and \sno{}, we include one Debye and one Einstein phonon, with parameters inspired by the DFT phonon dispersions, given in Figs.~\ref{fig:phonons+elph} in the main text and S2 in the SM~\cite{SM}. Slightly different Debye temperatures and speed of sound were required for a good fit using the Kukkonen and Bloch models, with the parameters for the latter in parentheses in the table. For \svo{}, an additional Einstein mode is included in the Kukkonen model. The diameter of the cylinder is approximated from the DFT Fermi surfaces, as shown in the insets in Fig.~\ref{fig:rho_kukkonen}. Overall, the main fitting parameters were the magnitudes of the deformation potentials.

The model fits are shown compared to the DFT+BTE calculations (reproduced from Fig.~\ref{fig:resistivity} in the main text) in Fig.~\ref{fig:rho_kukkonen}, including the separate contributions from the Debye and Einstein modes. Both models can fit the low-$T$ $T^5$ and high-$T$ $T$-linear regime (the latter with the help of the Einstein mode) but the Bloch model fails to capture the intermediate $T^2$ regime in \smo{}, \swo{}, \sto{}, and \sno{}. This indicates that this regime is indeed caused by the cylindrical Fermi surface, especially because the trend in the $T$ range of $T^2$ matches the DFT Fermi surfaces (see insets).

\svo{}, however, shows somewhat different behavior. For one, the $T^2$ region is at higher $T$ than would be expected from the radius of the Fermi-surface cylinder. Also, an additional Einstein mode is required for the Kukkonen model to fit the DFT+BTE data, and the energy of the Einstein mode(s) is much lower than in the other materials, consistent with the phonon dispersion (Fig.~\ref{fig:phonons+elph}). In the end, the Bloch and Kukkonen models fit equally well, indicating that the mechanism for $T^2$ is different in this case, and it originates from the thermal activation of optical modes as argued in Ref.~\citenum{abramovitch2024}. This behavior is consistent with the picture of higher RT resistivity in \svo{} originating from the lower energy of the optical modes, which complicate the $T$ dependencies predicted by either the Bloch or Kukkonen model. 

\end{document}